\documentstyle[12pt]{article}

\newcommand{\be}{\begin{equation}}
\newcommand{\ee}{\end{equation}}
\newcommand{\bear}{\begin{eqnarray}}
\newcommand{\ear}{\end{eqnarray}}

\date{}
\begin{document}
\begin{titlepage}
\begin{flushright}
HD--THEP--94--23\\
OUTP--94--12 P
\end{flushright}
\quad\\
\vspace{1cm}
\begin{center}
{\bf\LARGE NON-PERTURBATIVE ANALYSIS}\\
\medskip
{\bf\LARGE OF THE COLEMAN-WEINBERG}\\
\medskip
{\bf\LARGE PHASE TRANSITION}\\
\vspace{1cm}
D. Litim\footnote{e-mail: CU9@IX.URZ.UNI-HEIDELBERG.DE}
and C. Wetterich\footnote{e-mail: WETTERIC@
POST.THPHYS.UNI-HEIDELBERG.DE}\\
Institut  f\"ur Theoretische Physik\\
Universit\"at Heidelberg\\
Philosophenweg 16, D-69120 Heidelberg\\
\bigskip
N. Tetradis\footnote{e-mail: TETRADIS@THPHYS.OX.AC.UK}\\
Department of Theoretical Physics\\
1 Keble Road, Oxford OX 3NP, U.K.\\
\vspace{1cm}
\abstract{
We perform a non-perturbative study of the Coleman-Weinberg phase
transition in scalar QED. Our method permits a consistent treatment
of the effective potential near the origin, a region not accessible to
perturbation theory. As a result, we establish reliably the first order
character of the phase transition for arbitrary values of the scalar
coupling. }
\end{center}

\end{titlepage}
\newpage
The seminal Coleman-Weinberg one-loop computation \cite{1} of the
effective
potential in scalar quantum electrodynamics revealed the phenomenon
of spontaneous symmetry breaking induced by quantum fluctuations.
This perturbative calculation
indicates a first order transition between the symmetric phase and
the one with spontaneous symmetry breaking in dependence on the
scalar mass term. This observation has motivated
many theoretical investigations, ranging from a lower bound on the
Higgs
mass in the standard model (for top quark mass below 80 GeV) -- the
Linde-Weinberg bound \cite{2} -- to speculations about the
first order
character of the electroweak phase transition at high temperature
\cite{3}.
The latter could have important consequences for the creation of the
baryon asymmetry in the early universe \cite{4} by providing a new
scenario for non-equilibrium conditions at temperatures of the order
of the Fermi scale.
Another area of applications are models of inflationary cosmology
where
Coleman-Weinberg type potentials are frequently used \cite{5}.

Strictly speaking, however, the perturbative calculation is only
valid
for the convex part of the effective scalar potential where all mass
terms for
fluctuations are positive or zero and the saddlepoint expansion is
well
defined. The failure of the loop expansion for negative mass terms
for the
fluctuations manifests itself through an imaginary part in the effective
potential. The region in field space relevant for the first order
character of
the phase transition does not belong to the convex part of the
potential
and is therefore not directly accessible to perturbation theory.
Since the
problems are related to the fluctuations of scalar fields and the
scalar
interactions are small in the critical region, one usually discards
the
scalar fluctuations and only considers contributions from
gauge fields.
Indeed, the quartic scalar coupling $\lambda$ is much smaller than
$e^4$
(with $e$ the abelian gauge coupling) in the critical region and the
neglection of scalars seems justified at first sight. Unfortunately,
this
assumption cannot be verified within the standard perturbative
framework
since a saddle point expansion is crucial for its validity.
Furthermore,
little is known about the effect of higher-order loop corrections on
the
weak first order character of the transition.

Recently, a new non-perturbative method for the study of spontaneous
symmetry breaking has been proposed. It is based on the concept of
the average
action \cite{6} which is the continuum analogue of the blockspin
action
\cite{7} in lattice theories. The average action $\Gamma_k$ is the
effective action for averages of fields taken over a (Euclidean)
volume $\sim k^{-4}$.
It is obtained by integrating out the quantum fluctuations with momenta
$q^2>k^2$. The dependence of $\Gamma_k$ on the ``average scale'' $k$
is governed by an exact non-perturbative evolution equation \cite{8}
\footnote{See ref. \cite{8a} for earlier versions of exact
renormalization
group equations.}.
It can be shown \cite{8} that the effective average action
interpolates
between the classical action as defined at some short distance scale
$\Lambda^{-1}$ $(\Gamma_\Lambda\equiv S)$ and the generating
functional of the 1PI Green functions -- the usual effective action
$\Gamma$ -- for $k=0$
$(\Gamma_0\equiv \Gamma)$. The properties of the phase transition can
be read
off from the average potential $U_k$ which is a manifestly real
quantity
for all values of the scalar field. Even for regions where
perturbation
theory ceases to be valid the non-perturbative evolution equation --
which
takes the form of a renormalization group-improved one-loop flow
equation --
still applies. This method seems therefore suitable for an
investigation
of non-perturbative aspects of the Coleman-Weinberg phase transition.

The concept of the average action was generalized for abelian
gauge
theories \cite{9}, \cite{10}. The dependence of the effective scalar
potential
$U_k$ \cite{9} and the gauge coupling \cite{10} on the average scale
$k$ was computed in arbitrary dimensions, demonstrating how the new
non-perturbative method can successfully cope with the infrared
divergences
present in the standard perturbative approach. The exact
non-perturbative
evolution equation for the scale dependence of the effective average
action
has been adapted to non-abelian \cite{11} and abelian \cite{12} gauge
theories\footnote{See ref. \cite{12a} for
 a different approach.}. It appeared
that the flow equations for the scalar potential
and the gauge coupling proposed in \cite{9},\cite{10} coincide with
the exact non-perturbative evolution equation in the limit where the
effective action is truncated to include only standard kinetic terms
(with wave function renormalization) for the scalar and gauge fields
as well as a scalar potential.

In this letter we employ the evolution equations for a
non-perturbative
study of the phase transition of the abelian Higgs model. We restrict
the
discussion to four dimensions and vanishing temperature and study the
phase
transition as a function of the mass term, as in the original papers
\cite{1},
\cite{2}. We are mainly interested in the phase structure of the
theory
and concentrate on the scale-dependent effective scalar potential
$U_k(\rho)$,
where $\rho=\varphi^*\varphi$. The exact evolution equations for
$U_k$ can
be infered from ref. \cite{12}, but are far too complicated to be
solved
exactly. We therefore proceed to different approximations -- all of
them
non-perturbative and going substantially beyond the one-loop
computation. We first consider  the limit where the scalar
fluctuations can be neglected
and deal with the scalar fluctuations later.
Furthermore, we approximate the derivative terms in $\Gamma_k$ by
standard kinetic terms for both the scalar field and the gauge field
and neglect the wave function renormalization of the scalar field.
The flow equation for the scale dependence of the average potential
becomes
very simple in this case and reads, with $t=\ln (k/\Lambda)$,
\be\label{U1}
\partial_tU_k''(\rho)=-\frac{3}{8\pi^2}e^4\int^\infty_0dx\
x\partial_t(P(x)
+2e^2\rho)^{-2}.\ee
Here primes denote derivatives with respect to $\rho$, $e^2(k)$ is
the running abelian charge, and the effective inverse propagator
\be\label{U1'}
P(x)=x\left[ 1-\exp\left(-\frac{x}{k^2}\right)\right]^{-1}\ee
is a function of the momentum squared $q_\mu q^\mu=x$, which contains an
effective
infrared cutoff $P(0)=k^2$. The partial derivative $\partial_t$ on
the r.h.s.
of eq.~(\ref{U1}) is meant to act only on $P$.

After performing the momentum integration the flow equation
(\ref{U1})
becomes a partial differential equation for $U_k$ as a function of
the two
variables $\rho$ and $k$. It has to be supplemented by an equation
for the
running of the abelian gauge coupling $e^2(k)$. Since the latter is
only
logarithmic -- the usual one-loop $\beta$
function is a good approximation (cf.~eq.~(\ref{E1})) -- we may first
simplify
for constant $e^2$. The evolution equation (\ref{U1}) can then be
solved
easily in the form
\be\label{U2}
U_k'''(\rho)=\frac{3}{2\pi^2}e^6\int^\infty_0 dx\
x(P+2e^2\rho)^{-3}.\ee
Taking the limit $k\to0$, we obtain the effective potential (apart
from
an irrelevant constant)
\be\label{U3}
U_0(\rho)=m^2_s\rho+\frac{3}{16\pi^2}e^4\rho^2\left[
\ln\left(\frac{\rho}{\rho_0}\right)
-\frac{1}{2}\right],\ee
where $\rho_0$ is fixed through $U_0'(\rho_0)=0$. This is exactly the
well-known
one-loop result of Coleman and Weinberg \cite{1}. The mass term at
the
origin $m^2_s$ is obtained as an integration constant and may hence be
treated
as a free parameter. The second free integration constant corresponds
to the quartic scalar coupling taken at some appropriate scale, i.e.~
$U''(\rho=\mu^2)$, or, equivalently, the scale $\rho_0$. As a
function
of $m^2_s$ the theory indeed exhibits a first order phase transition
in
this approximation!

Beyond the approximation of constant $e^2$ the logarithmic running of
$e^2(k)$
does not alter the qualitative features of the phase transition. A
numerical
solution of eq.~(\ref{U1}) with running $e^2(k)$ gives only minor
quantitative
corrections. This also holds if the gauge boson contribution to the
wave function renormalization (anomalous dimension) of the scalar
field is included (see below). The only remaining difference between
the exact
evolution equation and the flow equation in the limit of neglected
scalar
fluctuations concerns then a generalization of the gauge boson
kinetic term
in $\Gamma_k$ beyond the standard form $\sim F_{\mu\nu}F^{\mu\nu}$
and a similar generalization of the covariant scalar kinetic term.
These effects would modify the truncated form $(P(x)+2e^2\rho)^{-1}$
of the effective gauge boson propagator in the
presence of a constant background scalar field $\rho$.
In principle, the scale dependence of these modifications can in turn
be determined from exact flow equations. In the limit of small gauge
couplings they could alternatively be approximated by one-loop
expressions
-- this would allow a determination of the $e^6$-terms in the
$\beta$-functions
without ever doing a two-loop calculation, in complete analogy to the
successful demonstration in pure scalar theories \cite{12b}.
It is easy to convince oneself that none of these small propagator
modifications could possibly alter the first order character of the
phase transition
indicated by the flow equation (1). The dominant effect  $-\
U_k''(\rho)$ turning negative for small enough $\rho$ as $k$ decreases
-- is independent
of the details of the effective propagator. We conclude that in the limit
of neglected scalar fluctuations the form of the average potential
$U_k$ indeed
indicates a first order phase transition. For small $e^2$ the
Coleman-Weinberg
potential (\ref{U3}) gives an accurate quantitative description, with
only
minor modifications $\sim e^6$ as compared to a more complete
non-perturbative
analysis.

The crucial question in order to firmly establish the first order
phase
transition concerns, however, the contributions to the effective
potential from fluctuations of the scalar field. Are they small as suggested
by
the small ratio $\lambda/e^2$? In order to give a non-perturbative
answer to this question we truncate the average action (with
background gauge fixing
\cite{12c})
\bear\label{X1}
\Gamma_k&=&\int d^dx\Bigl\lbrace
Z_{\varphi,k}(D^\mu\varphi)^*D_\mu\varphi+
U_k(\varphi^*\varphi)\nonumber\\
&&+\frac{1}{4}Z_{F,k}F_{\mu\nu}F^{\mu\nu}+\frac{1}{2\alpha}
(\partial_\mu(A^\mu-\bar A^\mu))^2\Bigr\rbrace.\ear
In this approximation and for $\alpha\to0$ the flow equation for the
average potential reads
\bear\label{X2}
\partial_tU_k(\rho)&=&3v_4\int^\infty_0dx\
x\partial_t\ln[P+2e^2\rho]\nonumber\\
&&+v_4\int^\infty_0dx\ x\partial_t\ln[P+U_k'(\rho)+2\rho U_k''(\rho)]
\nonumber\\
&&+v_4\int^\infty_0dx\ x\partial_t\ln[P+U_k'(\rho)],
\ear
with $v_4=1/32\pi^2$. As in (1), the partial derivative $\partial_t$
acts only on $P$.
We have solved this equation numerically and the corresponding phase
diagram is depicted in fig.~1. The basic variables needed in order
to describe the
shape of the average potential are the location of the minimum
$\rho_0(k)$ (where $U_k'(\rho_0(k))=0$) and the quartic coupling related to
$U_k''(\rho_0)$.
It is convenient to use dimensionless renormalized parameters, in
particular
$e^2(k)$, the renormalized abelian charge, $\lambda(k)$, the
renormalized quartic scalar self-coupling, and $\kappa(k)$, the
dimensionless renormalized
location of the potential minimum. Their definitions are as follows:
\be\label{D1}
e^2(k)=Z_{F,k}^{-1}\bar e^2(k)\ee
\be\label{D2}
\kappa(k)=k^{-2}Z_{\varphi,k}\rho_0(k)\ee
\be\label{D3}
\lambda(k)=Z_{\varphi,k}^{-2}U_k''(\rho_0),\ee
where $Z_{\varphi,k}(Z_{F,k})$ denotes the wavefunction renormalization
of the scalar (gauge) field.
In the projection of the phase diagram on the $\lambda,\kappa$ plane
we
observe a critical line separating the symmetric phase (SYM) from the
phase with spontaneous symmetry breaking (SSB). The other curves
represent the flow of $\kappa(k)$ and $\lambda(k)$ for initial values
$\kappa(\Lambda)$
and $\lambda(\Lambda)$ given at a fixed short distance scale
$\Lambda$.
(We use $e^2(\Lambda)=0.1$ .) The arrows represent the direction of the
flow with $k\to0$. Trajectories starting below the critical line end
with
$\kappa=0$ and, therefore, with the potential minimum at the origin. For
trajectories above the critical line $\kappa$ finally diverges $\sim
k^{-2}$ such that $\rho_0$ reaches a constant value corresponding to
the square of
the vacuum expectation value of $\varphi$.

For $\lambda>10^{-3}$ all quantitative details of the numerical
solution can be understood by a simple polynomial expansion of $U_k$
around
$\rho_0(k)$, setting $U_k'''(\rho_0)=0$ and similarly for higher
derivatives. In this approximation the evolution equations for the
SSB regime
read
\be\label{E1}
\partial_te^2=\frac{1}{24\pi^2}e^4\tilde
s^4_g(2\lambda\kappa,2e^2\kappa)\ee
\bear\label{E2}
\partial_t\lambda&=&\frac{1}{16\pi^2}\lambda^2(9s^4_2(2\lambda\kappa)+
1)
+\frac{3}{4\pi^2}e^4s^4_2(2e^2\kappa)\nonumber\\
&&-\frac{3}{4\pi^2}e^2\lambda
s^4_{1,1}(2e^2\kappa,2\lambda\kappa)+\frac{1}{4\pi^2}\lambda^3\kappa
m^4_{2,2}(2\lambda\kappa,0)\ear
\bear\label{E3}
\partial_t\kappa&=&-2\kappa+\frac{1}{16\pi^2}(3s^4_1(2\lambda\kappa)+1
)
+\frac{3}{8\pi^2}\frac{e^2}{\lambda}s^4_1(2e^2\kappa)\nonumber\\
&&+\frac{3}{8\pi^2}e^2\kappa
s^4_{1,1}(2e^2\kappa,2\lambda\kappa)-\frac{1}{8\pi^2}
\lambda^2\kappa^2m^4_{2,2}(2\lambda\kappa,0).\ear
(The last two terms in eqs.~(\ref{E2}) and (\ref{E3}) arise
from the anomalous dimension $\eta_\varphi=-\partial_t\ln
Z_{k,\varphi}$
and correspond to $2\eta_\varphi\lambda$ and $-\eta_\varphi\kappa$,
respectively.) The new elements, compared to the standard perturbative
one-loop $\beta$-functions, are the threshold functions $\tilde
s_g^4,s^4_1,  s^4_2, s^{\ 4}_{1,1}$ and $m^{\ 4}_{2,2}$. They
describe
the effective decoupling of particles with mass much larger than $k$.
Indeed, two different masses play a role in the regime with
spontaneous
symmetry breaking (SSB regime). The mass of the gauge boson at the
scale
$k$ is given by
\bear\label{Mone}
M^2(k)&=&2\bar e^2(k)\rho_0(k)Z_\varphi Z_F^{-1}\nonumber\\
&=&2e^2(k)\kappa(k)k^2\ear
whereas the mass of the Higgs boson corresponds to
\bear\label{M2}
m^2(k)&=&2\rho_0(k)U_k''(\rho_0)Z^{-1}_\varphi\nonumber\\
&=&2\lambda(k)\kappa(k)k^2.\ear
The physical masses of these particles (as defined by the two-point
function at zero momentum) are obtained from eqs.~(\ref{Mone}),(\ref{M2}) in
the limit
$k\to0$. The threshold functions are all normalized to one
in the limit of vanishing arguments. For $\kappa\to0$ the evolution
equations (\ref{E1}),(\ref{E2}) therefore reduce to the standard
perturbative
renormalization group equations. On the other hand, the threshold
functions
vanish for large arguments $M^2/k^2$ or $m^2/k^2$. Their precise form
depends on the precise form of the infrared cutoff function $R_k(x)$
(for
details see ref. \cite{13}).
With
$R_k(x)=Z_{\varphi,k}x\exp(-\frac{x}{k^2})[1-\exp(-\frac{x}{k^2})]^{-1
}$
they read for large values of the argument $(w\gg1)$
\bear\label{S1} \cite{13}
s_n^4(w)&=&6n\zeta(3)w^{-(n+1)}\nonumber\\
&&-12n(n+1)[\zeta(3)+\zeta(4)]w^{-(n+2)}+{\cal O}(w^{-(n+3)})\ear
and similarly for the other threshold functions. The appearance of threshold
functions in the $\beta$-functions is physically expected. Particles with
masses much
larger than $k$ should not influence the running of couplings with
the infrared
cutoff scale $k$. Their mass already acts as an effective infrared
cutoff
and the variation of a second, much smaller cutoff $k$ should only
have
a small effect.

The second new ingredient is the evolution equation for $\kappa$. It
is closely
related to the ``quadratic renormalization'' \cite{6}
of $\rho_0$, or, equivalently, the scalar mass term. It describes
within a
renormalization group framework the physics related to the
``quadratic
divergences'' appearing in perturbation theory. Such an equation is
necessary in order to give a meaningful description of the threshold
effects
since those must depend on the ratio of the
renormalized ($k$-dependent) mass over $k$. Although well known in
practice in the Wilson approach \cite{7}
to renormalization (for example lattice studies), the effect of
quadratic
renormalization is missing in many versions of the renormalization
group
equations. This happens typically if there is no independent running
scale to which the masses can be compared, as for example in the
Coleman-Weinberg analysis of the effective potential were both the
gauge boson
mass and the running scale are given by the ``classical'' or
``background''
field $\varphi$.

In our approach the renormalization group equation for the
$k$-dependent
minimum of the average potential arises naturally from the minimum
condition
\be\label{Y}
\frac{d}{dt} U_k'(\rho_0(k))=\frac{\partial}{\partial t}U_k'(\rho_0)
+U_k''(\rho_0)\frac{\partial \rho_o}{\partial t}=0\ee
and the evolution equation for $\partial_tU_k'$
which involves terms $\sim e^2k^2$
and $\lambda k^2$. This
also explains the perhaps somewhat surprising negative power of
$\lambda$
in the third term of eq.~(\ref{E3}): For very small $\lambda$ even a
moderate
change in $U_k'(\partial_tU_k'\sim e^2k^2)$ induces a large change
in the location of the minimum.

Given a set of initial values
$e^2(\Lambda),\kappa(\Lambda),\lambda(\Lambda)$
at some short-distance scale $k=\Lambda$, the evolution equations
describe how the various couplings run in dependence on $k$. The
solution
for $k\to0$ gives the vacuum expectation value $\rho_0(k=0)$ as well
as the renormalized gauge coupling and quartic coupling at zero
momentum. The
value of $\rho_0(k=0)$ determines whether the theory is in the
symmetric phase $(\rho_0(0)=0)$ or in the SSB phase $(\rho_0(0)>0)$.
For a theory in the symmetric phase $\kappa$ reaches zero at some
scale $k_s>0$.
Once $\kappa(k_s)=0$, the evolution has to be continued using
equations appropriate for the symmetric regime.
Since most of the physical properties can be derived from the
equations in the
SSB regime, we will not present these equations explicitly here and
refer
the reader to ref. \cite{9}. For a theory in the SSB phase $\kappa$ remains
positive for
all values of $k$ and eqs.~(\ref{E1})-(\ref{E3}) are sufficient.
This system of coupled non-linear differential equations still looks
quite
complicated. It allows, however, for analytical solutions in some
approximations which make the main characteristics easy to
understand.

First we discuss the linear regime defined by
$\label{L0}
2\lambda\kappa\ll1, \ 2e^2\kappa\ll1$.
All masses are then much smaller than the scale $k$ and we should
recover
the results of perturbation theory. The evolution equations simplify
considerably:
\be\label{L1}
\partial_te^2=\frac{4}{3}v_4e^4\ee
\be\label{L2}
\partial_t\lambda=24v_4\left[\frac{5}{6}\lambda^2-\lambda e^2+e^4\right]\ee
\be\label{L3}
\partial_t\kappa=-2\kappa+8v_4-12v_4\lambda\kappa+12v_4\frac{e^2}{\lambda}
-24v_4\frac{e^4\kappa}{\lambda}+12v_4e^2\kappa.\ee
Equations (\ref{L1}), (\ref{L2}) coincide with the standard one-loop
result in perturbation theory. The third and fifth term of eq.~(\ref{L3}) are
obtained through the expansion of $s^4_1(w)$ for small $w$:
$s^4_1(w)=1-w+{\cal O}(w^2)$.
Combining eqs.~(\ref{L2}) and (\ref{L3}) one can obtain the
anomalous mass dimension of the scalar field in one-loop order.
The running of the gauge coupling
\be\label{L3'''}
e^2(k)=\frac{e^2(\Lambda)}{1+\frac{4}{3}v_4e^2(\Lambda)\ln\frac{\Lambda}{k}}
\ee
is a small logarithmic effect and will be neglected in the following
($e^2(k)=e^2$). There is no fixpoint for the running of $\lambda$ (or
for the evolution of the ratio $\lambda/e^2$) and the positive
$\beta$ function
drives $\lambda$ always towards zero. The interesting physics which
is associated with the first order character of the
Coleman-Weinberg-type phase transition occurs for values of the
quartic
coupling $\lambda$ much smaller than $e^2$, typically $\lambda\sim
v_4e^4$.
For $\lambda\ll e^2$ we can neglect the terms
$\sim\lambda^2$ and $e^2\lambda$ in eq.~(\ref{L2}). The running of $\lambda$
is then easily found
\be\label{L5a}
\lambda(k)=24v_4e^4\ln\left(\frac{k}{k_c}\right)\ee
\be\label{L5b}
k_c=\Lambda\exp\left(-\frac{\lambda(\Lambda)}{24v_4e^4}\right).\ee
According to eq.~(\ref{L2}) $\lambda$ reaches zero at some scale
$k_c\not=0$ and
becomes negative for $k<k_c$. To solve eq.~(\ref{L3}) for $\kappa$
in the limit $|\lambda|\ll e^2$ we observe the existence of an
UV-stable
fixpoint for $\lambda\kappa$:
\be\label{L6a}
\partial_t(\lambda\kappa)=-2[\lambda\kappa-(\lambda\kappa)_*]\ee
\be\label{L6b}
(\lambda\kappa)_*=6v_4e^2\ee
This fixpoint is represented by the dashed line in the phase diagram
in fig.~1. Combining the solution of eq.~(\ref{L6a}) with eq.~(\ref{L5a}) we
find
\bear\label{L7}
\kappa(k)&=&\frac{1}{4e^2}\frac{1+\Delta_\Lambda\frac{\Lambda^2}{k^2}}
{\ln (k/k_c)}\nonumber\\
\Delta_\Lambda&=&\frac{\lambda(\Lambda)\kappa(\Lambda)}{(\lambda\kappa
)_*}-1 .
\ear
We have parametrized the initial value at $\Lambda$ such that
$\Delta_\Lambda$
measures the deviation of $\lambda\kappa$ from its fixpoint value
of eq.~(\ref{L6b})
at the scale $\Lambda$.

Due to the running of $\lambda$ there is no fixpoint in $\kappa$. For
$\Delta_\Lambda=0$ one sees that $\kappa$ diverges for $k\to k_c$. On
the
other hand, for sufficiently negative $\Delta_\Lambda$, $\kappa(k)$
will
vanish at some scale $k_s>k_c$ and the evolution will continue in
the symmetric regime.
For $k<k_s$ we replace the parameter $\kappa(k)$ by
\be\label{MO}
m^2_s(k)=Z^{-1}_{\varphi,k}U_k'(0)\ee
and use evolution equations adapted to the symmetric regime. In
particular, the scale dependence of the mass term simplifies for
$|\lambda|\ll e^2$ to
\be\label{M1}
\partial_t m^2_s=-\frac{3}{8\pi^2}e^2k^2.\ee
The mass term at $k=0$ is therefore connected to $k_s$ by
\be\label{L9'}
m^2_s=\frac{3}{16\pi^2}e^2k^2_s.\ee
There is a critical value
\be\label{L8}
\Delta_{\Lambda_c}=-\frac{k^2_c}{\Lambda^2}\ee
separating these two different behaviours.
For $\Delta_\Lambda>\Delta_{\Lambda_c}$ we see that $\kappa(k\to
k_c)$
diverges, whereas for $\Delta_\Lambda<\Delta_{\Lambda_c}$ we end in
the
symmetric phase. For $\Delta_\Lambda=\Delta_{\Lambda_c}$ we find that
$\kappa(k\to k_c)$ reaches a finite non-zero value
\be\label{L9}
\kappa_c=\frac{1}{2e^2}.\ee
Thus the system of equations (\ref{L1})-(\ref{L3}) exhibits a
bifurcation phenomenon at the critical value $\Delta_{\Lambda c}$: No
finite non-vanishing
value of $\kappa(k_c)$ except $1/2e^2$ can be reached.

In the region of initial values $\Delta_\Lambda<2\Delta_{\Lambda_c}$
the linear
approximation of eqs.~(\ref{L1})-(\ref{L3}) remains always valid. In this
region
the mass term is bounded from below
\be\label{L9''}
m^2_s>\frac{3}{8\pi^2}e^2k^2_c=\frac{3}{8\pi^2}e^2\Lambda^2\exp\left(
-\frac{\lambda(\Lambda)}{24v_4e^4}\right).\ee
In the SSB phase $\kappa$ always diverges and there must be some
regime
where the linear approximation breaks down. A quick inspection
of the full equations (\ref{E2})(\ref{E3}) shows, however, that for
$\lambda>\frac{3}{8\pi^2}e^4$ the threshold functions only imply a
stop
of the running of $\lambda$ and $\rho_0$ for $2e^2\kappa>1$. We can
therefore approximate
\be\label{L9'''}
\begin{array}{cc}\partial_t\lambda=0\\
\partial_t\rho_0=0\end{array} \Biggr\rbrace\quad{\rm for}\quad
\Biggl\lbrace\begin{array}{cc}
2e^2\kappa>1\\
\lambda>\frac{3}{8\pi^2}e^4.\end{array}\ee
For $\Delta_\Lambda>0$ we find that $2e^2\kappa$ becomes larger than
one
at a scale $k_F$ where
$\ln(k^2_F/k^2_c)>1,\lambda(k_F)>\frac{3}{8\pi^2}e^4$.
For this region there is a lower bound on the vacuum expectation
value and the
Higgs mass
\bear\label{L94}
&&\rho_0(k=0)>\frac{\exp(1)}{2e^2}k^2_c=\frac{\exp(1)}{2e^2}\Lambda^2\exp\left
(-\frac{\lambda(\Lambda)}{24v_4e^4}\right)\nonumber\\
&&m^2>\frac{3\exp(1)}{8\pi^2}e^2k^2_c.\ear

In order to make a quantitative comparison with the perturbative
Coleman-Weinberg analysis \cite{1} for $\lambda={\cal O}(e^4)$ we
want
to identify the value $\kappa(\Lambda)$ which leads to the ``massless
theory'', i.e.~$m^2_s(k=0)=0$. For a quartic potential at the short
distance scale $\Lambda$ (and $Z_{\varphi,\Lambda}=1$) we observe
\be\label{L10}
m^2_s(\Lambda)=-\lambda(\Lambda)\kappa(\Lambda)\Lambda^2.\ee
Using the flow equations (\ref{L6a}) and (\ref{M1})
allows us to express the initial deviation $\Delta_\Lambda$ from
the $(\lambda \kappa)$-fixpoint (24)
in terms of the mass term at the origin $m_s^2$
for $k=0$:
\be\label{L11}
\Delta_\Lambda=-\frac{1}{6v_4e^2}\frac{m^2_s(k=0)}{\Lambda^2}.\ee
Therefore, the ``massless theory'' corresponds to $\Delta_\Lambda=0$.
Using the
linear equations (\ref{L5a}),(\ref{L7}) as long as $2e^2\kappa<1$ and eq.~
(\ref{L9'''})
for $2e^2\kappa>1$ we deduce for $m^2_s=0$ the mass relation (for
$k=0$)
\be\label{L12}
\frac{m^2}{M^2}=\frac{\lambda(k_b)}{e^2}=\frac{3}{8\pi^2}e^2 ,\ee
where $k_b$ is determined by $2e^2\kappa(k_b)=1$. Thus, in this
approximation
the well-known result of Coleman and Weinberg \cite{1} is recovered.

We conclude that except for the immediate vicinity of
$\Delta_{\Lambda_ c}(2\Delta_{\Lambda_c}<\Delta_\Lambda<0)$ the
system of equations (\ref{E1})-(\ref{E3}) reproduces the results of
the Coleman-Weinberg analysis, including the effect
of dimensional transmutation (cf.~eqs.~(\ref{L9''}),(\ref{L94})
and (\ref{L12})).
It is tempting
to think that the bifurcation phenomenon observed in the linear
approximation
corresponds to a first order transition, leading to mass bounds
somewhat
weaker than (\ref{L9''}) and (\ref{L94}). We observe that the bifurcation
occurs for
values $\lambda\ll e^4$. If the effect of scalar fluctuations could
be neglected for very small $\lambda$, we may use eq.~(\ref{U1}) to
establish
the first order character of the phase transition. Using the system
of
flow equations (\ref{E1})-(\ref{E3}) for $\lambda>e^4$ and eq.~(\ref{U1})
for $\lambda<e^4$
 would then give a detailed non-perturbative
quantitative picture of the Coleman-Weinberg phase
transition\footnote{The two systems of flow equations have a smooth
overlap for small $e^2$.}.

For a numerical study of the evolution equation (\ref{X2}) for
$\lambda\ll e^4$ we have to consider potentials which may have more
than one local minimum.
For this reason we use for our numerical investigation an
approximation
of $U_k$ as a quartic polynomial in $\rho$, i.e.~we include couplings
up to $\varphi^8$. As long as $U_k$ exhibits a local ``asymmetric
minimum''
$(\lambda>0,\kappa>0)$, the effective average potential is
parametrized
in terms of the four variables $\lambda,\kappa, m_s^2/k^2, $ and
$\lambda_s$, where $\lambda_s(k)=Z_{\varphi,k}^{-2}U_k''(\rho=0)$ denotes
the renormalized quartic scalar self-coupling at the origin.
This parametrization is best suited for the study of a (weak)
first order
phase transition: We expect that the system near the phase transition
develops a small (positive)
mass term at the origin. Therefore, it is most
important that all the relevant quantities like masses and quartic
couplings are followed by their appropriate evolution equation.
Higher
couplings like $U_k'''(\rho_0)$ appearing on the r.h.s. of the
evolution
equation for $\kappa$ and $\lambda$ are determined as functions of
the four
variables using the polynomial approximation for $U_k$. Our
parametrization
yields information on the global structure of the effective potential
since it contains information on the symmetric extremum
$(m^2_s,\lambda_s)$
as well as on the asymmetric one $(\kappa,\lambda)$. When the broken
phase
disappears, ($\lambda\leq0$ or $\kappa=0$), the
evolution is continued in the symmetric regime with the corresponding
evolution equations. A more detailed account on this method and some
generalisations will be given elsewhere \cite{14}.

The phase diagram in the range of
 small values of $\lambda(k)$ is shown in fig.~2. We have
specified the initial conditions at some fixed short distance scale
$\Lambda$ by the requirement that the classical potential $U_\Lambda$
is quartic in the fields,  $U(\rho)=\frac{1}{2}\lambda(\Lambda)(\rho
-\rho_0(\Lambda))^2$. For all trajectories, we fix $e^2(\Lambda)$
and $\lambda(\Lambda)$. The phase diagram is then obtained through the
variation
of $\kappa(\Lambda)$ which serves as a free parameter.

The phase transition is clearly of the weak first order type and we
observe three characteristic regions in the $(\lambda,\kappa)$ plane:
In region I the
average potential has only one minimum. (This means $m^2_s<0$ for
$\kappa>0$.)
For trajectories remaining within region I this holds for all values
of $k$. Once $\kappa(k_s)$ becomes zero, region I should be continued
in the symmetric regime with the only minimum now at the origin. The
behaviour
of trajectories remaining within region I is very similar to a
second order
phase transition. In particular this concerns the values of $\lambda$
shown in fig.~1. For not too small values of $\lambda(\Lambda)$ the
phase
transition is therefore ``almost of the second order''. The weak
first order
character of the transition is related to trajectories so close to
the
critical line that they cross the dashed boundary between regions I
and II.
For regions II and III the average potential has two local minima.
The
``asymmetric minimum'' occurs for $\kappa>0$ and is the absolute
minimum
in region II. The ``symmetric minimum'' is at the origin. This
becomes the
lowest minimum in region III.
The (dashed) borderline separating I and II corresponds to $m_s^2=0$.
Therefore
the ``massless theory'' (the Coleman-Weinberg case) is described by
the trajectory A for which $m^2_s(k=0)=0$. On this trajectory the
mass relation
(\ref{L12}) is found to coincide with the numerical result within
4\%.
The regions II and III are separated by the dashed-dotted line. This
borderline
describes the effective potential with degenerate minima
$U_k(0)=U_k(\varphi_0)$. The trajectory with initial value
$\kappa_{crit}$ is
running asymptotically towards a  potential
with degenerate minima and corresponds to the
critical line of the phase diagram. All trajectories to the right of
B with
$\kappa(\Lambda)>\kappa_{crit}$ belong to the phase with spontaneous
symmetry breaking, whereas all trajectories to the left with
$\kappa(\Lambda)<\kappa_{crit}$ will end up for $k\to0$ with lowest
minimum at
the origin and therefore belong to the symmetric phase. The
trajectory
C is the spinodal line of the model. It describes, similarly to A, an
asymptotically massless theory but now for scalar fields at the
asymmetric local minimum. The first order part of the phase diagram
can only be seen
in the small interval of initial values $\kappa_\Lambda$ with
$\kappa_\Lambda^D<
\kappa_\Lambda<\kappa_\Lambda^A$. This corresponds to $\lambda$
smaller than
approximately $\frac{1}{32\pi^2}e^4$ in the final part of the running.

In conclusion, the weak first order character of the phase transition
has now been established on firm non-perturbative grounds.  In
particular, we emphasize that our non-perturbative evolution equation
(\ref{X2}) is valid
for all values of $\rho$, in contrast to the perturbative treatment
which
becomes unreliable around the origin, as pointed out in the original
work
\cite{1}. The average potential is always real. The perturbative
imaginary part
has to be considered as a pure artefact of perturbation theory. The
average
potential becomes convex for $k\to0$ as can be seen on very general
grounds
\cite{6},\cite{8}. This approach to convexity is described by the
exact evolution
equation, but any polynomial approximation for the potential breaks
down in the
SSB phase\footnote{For a discussion of the approach to convexity for
the
pure scalar theory see ref. \cite{13a}.}. We are not interested here
in the physics related to the approach to convexity and stop the
running of
$U_k$ once $k^2$ becomes of the order of the curvature of the
potential
at a local maximum. At these values of $k$ the masses $m^2_s(k),
m^2(k)$ or
$M^2(k)$ have already reached almost constant values (and similarly for
the
interactions), and we use in practice these values instead of the
values
for $k=0$. We can then proceed  to compute the physical photon mass
$M^2$
(which is proportional to the vacuum expectation value $\rho_0$), the
scalar
mass in the asymmetric minimum $m^2$, and at the origin $m^2_s$ as a
function
of the initial value $\kappa(\Lambda)$. As discussed before we may
use $m^2_s$
to label the distance from the phase transition. In fig.~3 we have
plotted
the various masses as a function of $m^2_s$ and indicated values
corresponding to the curves A, B and C in fig.~2. The mass ratio
$m^2/M^2$ is
given in fig.~4 together with the Coleman-Weinberg prediction
(\ref{L12})
for $m^2_s=0$. The visible small difference between our curve and the
prediction
(\ref{L12}) is actually not a real effect due to the scalar
fluctuations
neglected for eq.~(\ref{L12}) but rather an artefact of the polynomial
approximation to the potential. This can be seen by using the
numerical
solution with polynomial approximation, where the scalar fluctuations
are discarded. The resulting curve for $m^2/M^2$ is
essentially identical to fig.~4. We conclude that the effects of
scalar fluctuations are really
tiny for the small values of $\lambda$ relevant for the first order
character
of the transition. Our non-perturbative approach permits a very precise
quantiative
understanding of all physically interesting masses and couplings as
well
as of the complete shape of the effective potential. This is achieved through
the combination of the numerical solution of eq.~(\ref{X2}) in the range of
  $\lambda(k)>10e^4$  (cf.~fig.1) with
the analytical solution (\ref{U2}) for $\lambda(k)<10e^4$.

In summary, we find a first order phase transition for scalar QED for
arbitrary
values of the quartic scalar coupling $\lambda$. The mass gap between the
symmetric phase and the one with spontaneous symmetry breaking is
exponentially small (except for tiny values of $\lambda$). This is an aspect
of dimensional transmutation \cite{1} due to the logarithmic running of
$\lambda$. In the vicinity of the phase transition the running quartic
scalar coupling is always very small. As a result, we find that the one-loop
calculation of the effective
potential
in the abelian Higgs model \cite{1} gives a rather accurate
description
of all physical quantities of interest in the vicinity of the phase
transition.
Our non-perturbative method permits (for the first time) a consistent
treatment
of the region around $\varphi=0$ in the phase with
spontaneous
symmetry breaking. Comparison with our non-perturbative results
gives a simple prescription of how to deal, in this particular situation,
with the effects
of scalar fluctuations in the one-loop calculation:
The unphysical imaginary part should be discarded or the scalar fluctuations
should simply be
omitted, in the range of $\varphi$ for which a negative mass squared for the
scalar fluctuations  is predicted by the
one-loop calculation. The one-loop potential computed with this
prescription
is very close to the average potential $U_k$ at a scale $k_{cv}$
which is
given by $k^2_{cv}\approx 2m^2_{max}$, with
$m^2_{max}=-U'(\rho_{max})$
corresponding to the curvature at the local maximum of the potential.

More generally, the effective potential $U_0$ is not the most appropriate
quantity for a description of spontaneous symmetry breaking. On general
grounds $U_0$ is convex and therefore more subtle to handle in the phase with
spontaneous symmetry breaking. On the other hand, the average potential $U_k$
is not convex and the quantum effects leading
to convexity of the effective potential $U_0$ only set in for
$k<k_{cv}$ \cite{13a}. As long as the relevant physical distances are
smaller than
$k^{-1}_{cv}$ the fluctuations leading to convexity can be neglected.
This also holds for all properties related to the true vacuum
(corresponding to the absolute minimum of the potential), such as the masses
of physical particles and their interactions.
We should mention, however, that the flattening of the potential due to the
approach to convexity may play a role in certain cosmological scenarios of
inflation.

The present work has also permitted precise quantitative  tests for
analytical and numerical approximate solutions of the non-perturbative
evolution equations. These evolution equations remain valid in three
dimensions or for a
non-vanishing temperature where perturbation theory fails due to
severe
infrared problems. In the presence of large scalar couplings the one-loop
approximation neglecting scalar fields is not expected to give a
qualitatively correct description. The present work is therefore
an excellent starting
point for a non-perturbative treatment of the three-dimensional
abelian
Higgs model -- which should describe the phase transition of
superconductors
-- or of  the high temperature phase transition in  four-dimensional
scalar
quantum electrodynamics.

\vspace{1cm}
\section*{Figure Captions}
\vspace{.5cm}
\begin{description}
\item{Fig.~1:} The phase diagram of scalar QED. The thick
line
separates the symmetric phase (SYM) from
the phase with spontaneous symmetry breaking (SSB).
\item{Fig.~2:} The phase diagram for small $\lambda$. We indicate
the ``massless'' (A), critical (B), and spinodal (C)
trajectory.
\item{Fig.~3:} The mass of the gauge field M as a function of the scalar
mass term at the origin $m_s^2$. We also indicate the scalar mass $m$
for spontaneous symmetry breaking. All masses are given in units of the
cutoff $\Lambda$.
\item{Fig.~4:} The ratio of the Higgs-boson to gauge-boson mass as a function
of
m$^2_s$. The star denotes the Coleman-Weinberg result.
\end{description}
\end{document}